\documentstyle[11pt,paspconf]{article}

\begin{document}

\title{The cosmological parameters from supernovae\footnote{Lecture
given at the Casablanca Winter School 
``From Quantum Fluctuations to Cosmological 
Structures'', December 1996.
}}

\author{Pilar Ruiz--Lapuente}
\affil{Department of Astronomy, Faculty of Physics, U. Barcelona, Spain}
\affil{Max--Planck Institut f\"ur Astrophysik, Garching bei M\"unchen, Germany}

\begin{abstract}

\smallskip 

 Supernovae are bright luminous stellar objects observable up to 
 redshifts close to z $ \sim$ 1. They are used to probe 
 the geometry of the Universe and its expansion rate 
 by applying different methods. In this text, I review the 
 various approaches used to measure the present expansion rate of the 
 Universe,
 $H_{0}$, and the paths to determine its matter density 
 $\Omega_{M}$ and the possible contribution of a non--zero
 cosmological constant $\Lambda$. An account is given of
 the numerical estimates 
 of those cosmological parameters
 according to the present status of the research.

\end{abstract}


\section{Introduction}

 The quest for three numbers $H_{0}$ (the present value of the 
 Hubble parameter, i.e. Hubble constant), $\Omega_{M}$ 
 (the matter density of the Universe, i.e. density parameter)
  and $\Lambda$ (the possible non--zero
 value of the cosmological constant) has become 
  of primordial interest since the beginning of modern cosmology. 
  The type of Universe in which we live is described by those numbers,
  and its evolution is related to them. 

\bigskip

\noindent
  Original attempts to determine those values (see Weinberg 1973, for a review)
  involved the selection of a number of astrophysical objects which could
  be observed due to their brightness up to very large redshifts,
  and with properties of intrinsic luminosity well understood --i.e. 
  either homogenous in luminosity or following a well--known correlation
  with a measurable observable. 

\bigskip 

\noindent
   Supernovae (SNe) were depicted as one of those objects, and 
  their observed brightness along redshift space (also called Hubble diagram)
  suggested a way to trace the geometry and expansion of the Universe.

\bigskip

\noindent
  Along the following decades of research on these objects, very much 
  has been learnt about their physical nature and observational properties
  such as spectra and light curves. 
  We know that there are two ways in which a star can explode giving rise
  to what we know as a supernova. For massive stars the explosion occurs 
  through the gravitational collapse of their core at the end of their
  evolution. This class constitutes the ``gravitational 
  collapse SNe'' which includes the phenomenological Types II and Ibc
  (SNe II and SNe Ibc).
  A second way in which stars explode is through a thermonuclear 
  runaway of their degenerate core. This occurs for stars of low
  mass in binary systems which end up their lives as C+O degenerate cores: 
  white dwarfs (WDs). If those stars belong to a close binary system,
  and accrete matter from the companion, the growth in mass produces an 
  increase in density and temperature of the core which finally leads to
  the thermonuclear runaway. The ``thermonuclear supernovae'' correspond
  to the phenomenological class of Type Ia SNe (SNe Ia). 

\bigskip 

\noindent
   Being both SNe II and SNe Ia very bright events, a number of methods
   have been proposed to use their luminosity for cosmology purposes. 
   Some of them have a purely empirical basis and some others involve 
   a theoretical perspective. 
   SNe Ia are more homogeneous than SNe II. This placed them as more
   favorable objects to be used as ``standard candles'' for cosmological
   purposes.
   I will restrict the discussion to what 
   is being learnt through SNe Ia, actively used nowadays not 
   only for the $H_{0}$ determination, but for that of
   $\Omega_{M}$ and $\Lambda$ as well.     

\bigskip

\section{Dispersion and relationships of Type Ia supernova luminosity}

\bigskip

A standard candle to be used in the determination of the cosmological
parameters  should 
be able to depict the geometry of the Universe and its expansion rate
with as much accuracy as possible. Ideally such a precise indicator
would be a kind of ``Cepheid'' observable up to very large redshifts:
 an object whose magnitude could be predicted or 
known with very low intrinsic dispersion, its empirical 
relation with other observable properties being
 well understood. In Cepheids, the zero--point of 
the period--luminosity (P--L)
relationship is known with a 0.1 mag of accuracy, and the dependence of the 
P--L relation with metallicity has also been well studied and can not be
 larger than 0.05 mag (Feast \& Walker 1987). 

\bigskip

\noindent
 We do not have objects so well--known in magnitude as Cepheids,
 which could be discovered and observed at redshifts close to 
 z $\approx$ 1. However, Type Ia supernovae are likely, among the
 existing distance indicators, the closest one   
 to a ``long--distance Cepheid'' that we can find: 
 the understanding of their intrinsic
 magnitude, and its correlation with observable properties such as 
 the shape of the light curve and the spectral characteristics reduce
 the intrinsic dispersion in the predictable luminosity of those 
 candles to $\sigma = 0.2$. 

\bigskip 

\noindent
  The history of how the understanding of SNe Ia has evolved and 
 how the relationships between luminosity and other observables developed
 covers a long period of debates. Pskovski (1977, 1984) first suggested 
 the relationship between absolute magnitude at maximum 
and rate of decline of the
 light curve. His assertion was objected on the grounds that it could
 be an effect of different intrinsic reddening (Boisseau \& Wheeler 1991). 

\bigskip

\noindent
  The follow--up of SNe Ia  
  (Maza et al. 1994; Filippenko et al. 1992a,b; Leibundgut et al. 1992)
  confirmed that such an intrinsic dispersion
  in luminosity and decline rates of the light curve were real. 
  Extensive and accurate observations were collected by the CTIO
  group which allowed to build up a sort of ``period--luminosity''
  relationship for Type Ia supernovae: a mathematical expresion
  in which the intrinsic luminosity and the rate of decline of
  the light curve is established. Hamuy et al (1996a,b) expressed it as: 

\begin{equation}
M_{MAX} = a + b[\Delta m_{15}(b) - 1.1]
\end{equation}

\noindent
where a and b are constants and $\Delta m_{15}$ are the magnitudes 
decreased in 15 days after maximum. They found this relationship to be:

\begin{equation}
M_{B} = -19.25 + 0.78\ [\Delta m_{15}(B) - 1.1] + 5 \ log (H_{0} / 65) 
\end{equation}

\noindent
with an intrinsic dispersion of only $\sigma \sim 0.2$ mags.

\bigskip

\noindent
   Riess, Press \& Kirshner (1995a) give an alternative 
  way to express the intrinsic luminosity of Type Ia supernovae
  as related to the rate of decline of the light curve. They account
  for an overall shape parameter describing the evolution in luminosity
  before maximum to well past maximum. 
   Their parameterization (see Figure 1), and the one by Hamuy and
  collaborators (1995; 1996a,b) give a comparable scale of magnitudes
  for specific SNe Ia.  
  Riess, Press \& Kirshner (1995a,b; 1996) 
  have used as well the color light curve shapes to
 determine the reddening  affecting the observed luminosity of the supernova.

\bigskip

\bigskip

\noindent
  Another independent line of research on SNe Ia by Tammann \& Sandage
 (Tammann \& Sandage 1995; Tammann et al. 1997, Tammann 1996) 
  and  Branch \& collaborators (Branch et al. 1995) suggests that
  if restriction is made to those 
  SNe Ia which fulfill a number of observable requirements from the
  spectroscopic point of view (Nugent et al. 1995 ), 
  or from the colors, i.e. omitting objects redder than B--V = 0.2 
 (Tammann \& Sandage 1995),
  the  use of the decline--luminosity relationship is not necessary. 
  According to those authors, the concept of spectroscopically 
 ``normal SN Ia'', also called ``Branch--normal'' 
  SNe Ia  is a sharp enough guidance to the luminosity of SNe Ia. 
  The intrinsic dispersion of those  ``normal''  SNe Ia  would be
   of $\sigma$=0.3 mag.

\bigskip  

\noindent
    Even within the empirical use of Type Ia supernovae for 
   distance determinations, a spread of usages has proliferated.
   The final values 
   of H$_{0}$  by various authors are still different and will be 
   mentioned in section 6.

\bigskip

\noindent
   So far, we are discussing the prolific use of SNe Ia as
  candles from nearby supernovae samples (z $\sim 0.1$).  
   To derive  H$_{0}$ is suitable to use a sample at z below 0.3 since
   at larger redshifts the contribution from the deceleration term
   can not be neglected. SNe Ia at z $\ge$0.3 serve to determine 
   $\Omega_{M} $ and $\Omega_{\Lambda}$. Those uses
    will be addressed in section 7.

\bigskip 

\noindent
   In the following section, I would like to contrast the empirical 
   usage of Type Ia  
   supernovae with the theoretical expectations. 

\vfill\eject

\bigskip

\begin{figure}
\input epsf2
\bigskip
\bigskip
\bigskip
\epsfxsize=250pt
\epsfbox{ariesscontr2.epsi}
\caption{}
\end{figure}

\bigskip
\bigskip

\noindent
Figure 1.   The use of light curve shapes to calibrate SNe Ia luminosities,
from Riess, Press \& Kirshner (1996).

\vfill\eject

\section{Boundaries on $H_{0}$ from  WD explosions}

\smallskip

 The absolute magnitude of an exploded WD 
 has a limit established by the maximum mass of any WD which explodes: 
 the Chandrasekhar mass $M_{Ch} \approx 
1.38 M_{\odot} \ ( Y_{e}/ 0.5)^{2}$, where $Y_{e}$ is the number of
 electrons per nucleon.
 A WD accreting mass beyond the Chandrasekhar mass would
  either undergo a gravitational collapse forming a neutron star, or
  it would form a exploding Chandrasekhar mass object plus an envelope 
  of some mass around. The former would be an underluminous explosion,
  and the second would not imply significantly different
  absolute magnitudes than the bare exploded C+O WD. 

\bigskip 

\noindent
   In the thermonuclear explosion of a Chandrasekhar WD,
  the generated kinetic energy,
  $E_{kin}$ and the radioactive energy $E_{^{56}Ni}$ are linked.
  The larger the radioactive energy from $^{56}Ni$, providing the
  luminosity, the highest expansion velocities $E_{kin}$ have the ejecta.  
  A fast expanding ejecta traps less efficiently the radioactive
  energy: thus a self--constraining play on the final absolute magnitude is  
  obtained in explosions of different energies to favor a maximum absolute
  magnitude which a Type Ia explosion could achieve. 
  That limit corresponds to a final minimum $H_{0}$ of 
about 50 km s$^{-1}$ Mpc$^{-1}$.

\bigskip

\noindent
   An upper limit to the value that $H_{0}$ 
  is provided by the discussion of the minimum mass of a exploding WD, and
  if that could correspond to what we see as SNe Ia. The exploration of
  the range of possible exploding WDs by various proposed mechanisms    
  suggests that the range of what we can observe if a whole range of WDs  
  below the Chandrasekhar mass explode is much wider that the objects we
  actually identify as SNe Ia. This argument disfavors $H_{0}$ larger than
  75 km s$^{-1}$ Mpc$^{-1}$ (see sections 4 and 5).

\section{Theoretical uses of Type Ia supernova through light curves}

\smallskip

\noindent
 The first estimates and the use of theory of Type Ia supernovae
 to determine their absolute magnitude are found in Arnett, Branch \&
 Wheeler (1985) who made a prediction of that value from 
 the light curve of a exploded WD at the Chandrasekhar limit.  
  That evaluation was based in the power of $^{56}Ni$ and the trapping
  of its radioactive decay energy 
  to provide the peak of luminosity of a Type Ia supernova. 
  Soon a number of difficulties appeared related to this approach: 
  the calculation of the bolometric (overall) luminosity of a Type Ia
  supernova can be undertaken with moderate efforts. However, most of
  the observed light curves are given in broad--band filters: B, V, R 
  \& I. To make predictions of how the energy is distributed in the 
  different colors requires to make calculations taking into account
  detailed opacities and NLTE effects, among others. 
  One could address the problem trying to estimate a bolometric
  correction to transform the blue and visual light curves into bolometric
  light curves. Some authors have attempted this, but the bolometric 
  corrections  can easily become a bag of errors, if not sustained
  by light curve calculations, and the luminosity
  of SNe Ia go beyond any reasonable value.  

\bigskip

\noindent
   The problem of calculating the color light curves was addressed
   in a number of papers by H\"oflich and Khokhlov (1995). They predict
   B, V, R, I light curves for Type Ia supernovae, and attempt a simultaneous
   determination of reddening. 
  Chandrasekhar explosions with differences in burning propagation could, 
  according to their calculation, account for the intrinsic dispersion
  in SNe Ia light curves.  
 Assigning to each individual supernova a reduced set of Chandrasekhar
  explosion 
  realizations (in which differences arise from burning propagation
  or variation in envelope conditions), distances to those supernovae 
  are derived. 

\bigskip
         
\noindent                          
   A second caveat of the method of predicting the absolute magnitude
   of Type Ia supernovae is formulated in the question of what if the 
   models considered for Type Ia supernovae are not correct after all. 
   Arnett \& Livne (1995) addressed this issue by considering a different
   mechanism for the explosion of a WD as a Type Ia supernova: the edge--lit
   detonations below the Chandrasekhar mass, and predicting how those light 
   curves would be, and what would be the dispersion and absolute 
   magnitude reached in those explosions. They found that the absolute
   maximum--rate of decline relationship can be easily accounted for if
   WDs encompassing a wide range of masses explode below the Chandrasekhar 
   mass. 
        H\"oflich et al (1997), on the contrary, argued 
   against that possibility by
   pointing that those sub--Chandrasekhar
   explosions give too--blue light curves as compared 
   with observations. They favor the explanation of the intrinsic 
  dispersion of Type Ia supernovae by variations within 
  the Chandrasekhar--mass model (H\"oflich et al. 1997).

\bigskip
  
\noindent
     Recently, Pinto \& Eastman (1996) argue that a number of different 
   effects playing into the light curve physics  have not been 
   addressed so far in calculations. They suggest that
   Chandrasekhar models do not give the sort of correlation between 
   maximum brightness and rate of decline observed among SNe Ia.

\section{The absolute magnitude of Type Ia supernovae and the density
 diagnostics}

\smallskip

 A different approach towards the absolute magnitude of Type Ia 
 supernovae is linked to the possibility of obtaining density diagnostics
 of the object which explodes. This lead us to the discussion on the 
 mass of the object that we see reflected in light curves and spectra. 

\bigskip

\noindent
     Forbidden line emission is an excellent tracer of the density profile 
  of an exploded object. In particular, ratios of lines of [Fe$^{+}$] and  
  [Fe$^{++}$] inform us on the electron density of the exploded object.
   Density profiles, mass of the WD which explodes and luminosity 
  are linked properties of the explosion.
   The most dense the WD the most easily will trap  the $\gamma$--ray
   photons of the decays $^{56}Ni \rightarrow ^{56}Co \rightarrow ^{56}Fe$  
   which power the luminosity. 

\bigskip

\noindent
    Through the density diagnostics provided by the late emission of
   SNe Ia it is possible to conclude on the correct model of
   explosion, and its luminosity  (Ruiz--Lapuente 1996).
   The long--wavelength spectral 
   comparison provides also a way to 
   determine the reddening E(B-V) affecting the SN light (Ruiz--Lapuente \&
   Lucy 1992). 

\begin{figure}
\input epsf2
\epsfxsize=270pt
\epsfbox{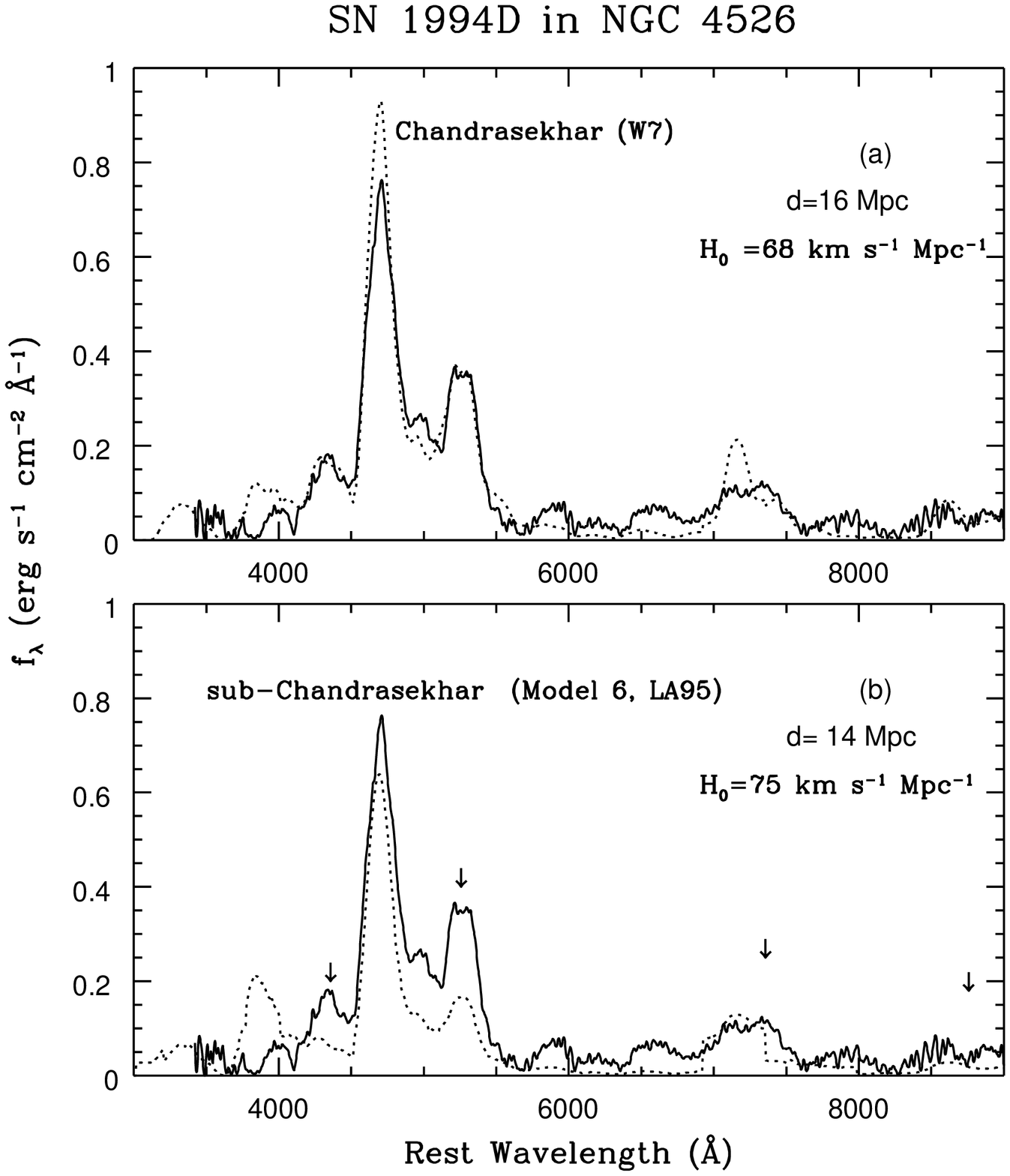}
\epsfxsize=200pt
\epsfbox{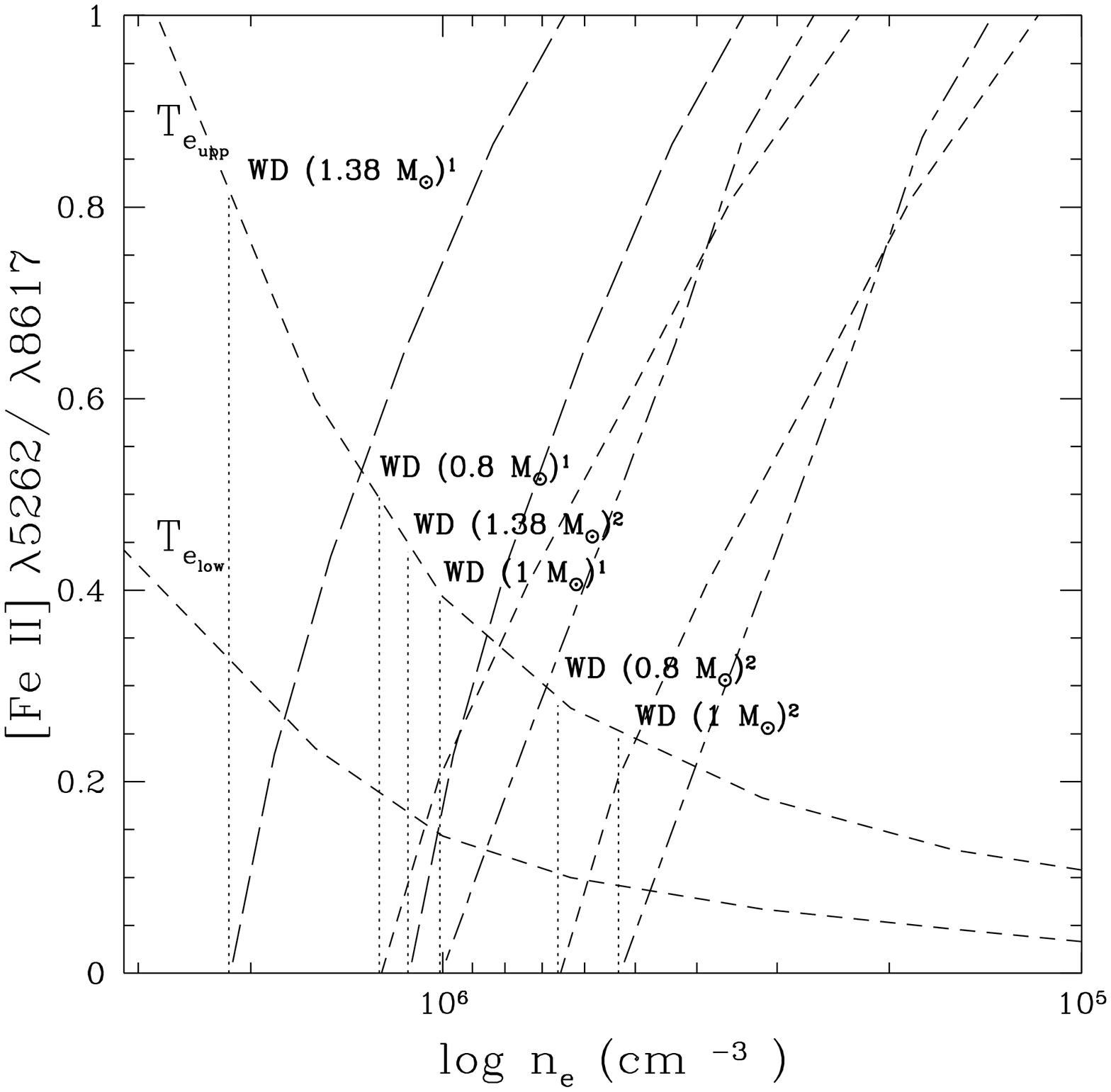}
\caption{ Top: Distance scale 
favored by sub--Chandrasekhar and Chandrasekhar 
 explosions and the discrimination through emission line.
  Bottom: Sensitivity of line emission to mass diagnostics.
  From Ruiz--Lapuente (1996).} 
\end{figure}

\bigskip

\noindent
  That sort of analysis suggests that explosions below the 
  Chandrasekhar mass (less dense explosions) fall outside the 
  n$_{e}$--T$_{e}$ plane (electron density and temperature)
   suggested by the observations of normal SNeIa. That produces 
  discrepancies between 
   the predicted relative luminosities of the lines and those displayed 
   by the SN.  The top panel of  Figure 2 shows the spectrum of
   a SN Ia from a Chandrasekhar WD at 300 days after explosion, 
   compared with a spectroscopically normal SNIa, and a comparison of
   the same SN with a sub--Chandrasekhar model. The bottom panel gives
   the sensitivity of the emission lines to electron density.

\section{Contrasting values of  $H_{0}$ from SNe Ia } 

\smallskip

 The initial discussion on the absolute magnitude of SNe Ia
 swept the wide range of values going from the mean value of
 absolute magnitude -18. to -20.5, having very different 
 consequences for the value
 of the Hubble constant.  The enormous progress made during the recent
 years has allowed to reduce considerably the uncertainty in that value.

\bigskip
 
\noindent
  From the empirical point of view the advance has come through the
 better study of the light curves of a large sample of supernovae and 
 from the possibility of using distances to individual galaxies 
 obtained through Cepheids to establish the zero point in the 
 calibration of the absolute magnitude of Type Ia SNe
 (Sandage et al. 1994).  From the theoretical
 point of view the improvement has come from the increasing sophistication in
 the calculations and the exploration of several mechanisms for the explosion
 of WDs. The path already covered has been amazing and a 
 convergence of views can be foreseen. 
    
\bigskip

\noindent
  To bring the story up to date, I will just compare some of the
 values preferred by the authors.

\bigskip

\noindent
   Tammann et al. (1996) have estimated the absolute
  magnitude of 7 SNe Ia from distances obtained from Cepheids 
  in the HST programme. They list mean absolute magnitudes 
  of  $ <M_{B} (max)> = -19.53 \pm 0.07$ and   
  $ <M_{V} (max)> = -19.49 \pm 0.05$ for ``spectroscopically 
   normal'' SNe Ia, and thereby $H_{0}= 56 \pm 3$ (Tammann et al. 
   1996).  Branch et al. (1997) formulate that for spectroscopically
   normal SNe Ia, (bluer colors than B--V=0.2) the empirical
   calibration gives:

\begin{equation}
 M_{B} \simeq M_{V} = -18.6 - 5 \ log(H_{0}/85)
\end{equation}
  
\noindent
  These values can be contrasted with theoretical
  approaches.  H\"oflich et al. (1997) obtain as a mean value for 
  the absolute magnitudes of SNe Ia: $<M_{B} (max)> = -19.40 \pm 0.2$ and 
  $<M_{V} (max)> = -19.37 \pm 0.18$.
  The absolute magnitudes from the forbidden emission approach using
  late--time SNe Ia spectra suggest a mean $ <M_{B} (max)> =-19.2 \pm 0.2$
  for ``spectroscopically normal'' SNe Ia and $ <M_{V} (max)> =-19.2 \pm 0.2$.
  Thus, a mean shift of 0.2 mag 
  in the central values compared with Tammann et al. (1996) results 
  (overluminous SNeIa such as SN 1991T would reach -19.5). 
  That 0.2 mag difference  shifts  from fifties to sixties
  the value of the Hubble constant.
  The method of light curves by H\"oflich et al. (1997) 
  favors  $H_{0}= 64 \pm 10$, and the late-emission method
  favors $H_{0}= 68 \pm 6 {\rm (stat)}
   \pm 7 {\rm (external)}$ (Ruiz--Lapuente 1996). Both approaches are 
  independent from the zero--point calibration from Cepheids. 

\bigskip

\noindent
  From the empirical approach, the
  relationship of absolute magnitude and rate of decline as derived
  by Riess, Press and Kirshner (1995a, 1996) and the zero--point calibration 
  from HST Cepheids, it is found that standard SNe Ia (shape parameter zero)
   have 
  $M_{V} (max) = -19.36 +0.1$. These authors obtain a value of 
   $H_{0} = 64 \pm  6 $.
  On his own, Hamuy et al. (1996a,b), using their parameterization of the
  rate of decline--brightness correlation 
  and a zero--point calibration from four supernovae (SM 1937C, SN 1972E, 
  SN 1981B, SN 1990N) out of the sample 
  of HST distance--calibrated SNe Ia,   
  obtain $H_{0}$ = 63.1 $\pm$ 3.4 (internal) $\pm$ 2.9 (external)
  km s $^{-1}$ Mpc $^{-1}$ and similar values for the 
  mean absolute magnitude of normal SNe Ia to Riess et al. (1996).
  Thus, theoretical methods and empirical ones taking into account
  the correlation of brightness and rate of decline suggest values  
  for the Hubble constant in the sixties range.

\bigskip

\noindent  
   The way from absolute magnitudes and individual distances to
   a global value of the Hubble constant presents divergences among authors. 
   It should be possible to establish the ``central'' value of the absolute
   magnitude of a SNIa: shape parameter zero for Riess et al. (1996);
   $\Delta m_{15} =1.1$ from Hamuy et al. (1996a,b); a given
   (B-V) color  according to Tammann et al. (1996) or spectral sequence 
   (Branch et al. 1996; 
   Nugent et al. 1996). Once the empirical correspondences are well 
   determined there should be little room left for a disagreement on
   $H_{0}$.

\bigskip
 
\noindent
    On the other hand, much confidence exists on the possibilities 
   of setting the final global value of H$_{0}$: 
   It has been shown that the value of
   H$_{0}$ provided by SNe Ia at high redshift should not differ significantly
   from the local value obtained with a nearby sample at z$\sim$ 0.1. 
   Kim et al (1996) find $ H_{0}^{L}$ / $H_{0}^{G}$ $<$ 1.10 at a 95$\%$
   confidence level.
   During the last years it has become more and more evident the depth of the
   Virgo cluster and the dangers of assuming a given galaxy to be at
   the core of that cluster. It has also been shown that  
   the path of using relative distances between Virgo and
   Coma  can be full of 
   errors (Tammann 1996; see also the article by Hendry in this volume). 
   These more controversial routes to the Hubble 
   constant  have not been used in the  
   the works mentioned in this section.
    Despite the increasing agreement on absolute magnitudes of SNe Ia,
   if compared with the starting point of that discussion, the 
   remaining difference 
   between a H$_{0}$ of 50--60 and 
   H$_{0}$ of 60--70 has important cosmological implications, and the
   discussion is not yet finished.

\section{From a higher redshift}

  Type Ia supernovae being among the brightest objects in the Universe,
 soon were used to determine
  the deceleration parameter $q_{0}$ which leads us to decide whether we 
  are in an open, flat or closed Universe.  The deceleration parameter
  $q_{0}$ measures the role of the matter density in slowing down the
  expansion rate of the Universe, and a possible contribution of a
  non--zero cosmological constant accelerating the expansion.

\begin{equation}
   q_{0} =  {1\over 2} \Omega_{M} - \Omega_{\Lambda}
\end{equation}

\noindent 
where  $\Omega_{\Lambda} =  \Lambda /3 {(H_{0})}^2$.
As it can be seen in equation 4, a negative $q_{0}$ means 
that $\Omega_{\Lambda}$ has a dominant contribution. 

\bigskip

\noindent
 The relationship of observed magnitude m with redshift z 
  for an object of constant luminosity has 
  a dependence with $q_{0}$ which 
 produces a bending in the  
  m(z) diagram (the Hubble diagram). That relationship departs from a 
 straight line at high z:

 \begin{equation}
 m = M - 5\ log H_{0} + 25 + 5\ log\ cz + 1.086\ (1 - q_{0})z + O(z^{2})
  \end{equation}

\noindent
   Soon, Oke \& Sandage  (1968) realized that a practical use 
  of that method implied to take into account the finiteness of the broad
  band filters in which one is collecting the light from distant objects. 
  As the light is redshifted towards longer wavelengths in an expanding
  Universe and the power of energy emitted by unit time is affected by
  the frequency shift resulting from the expansion, a correction from
  the measurement of the received luminosity
  through a filter at the site of emission
  and at a site of observation far from emission has to be included.  
  This correction, called K--correction, was pointed out by Oke \& Sandage 
  (1968) and has been applied thereafter (Kim et al. 1996).

\begin{figure}
\input epsf2
\epsfxsize=250pt
\epsfbox{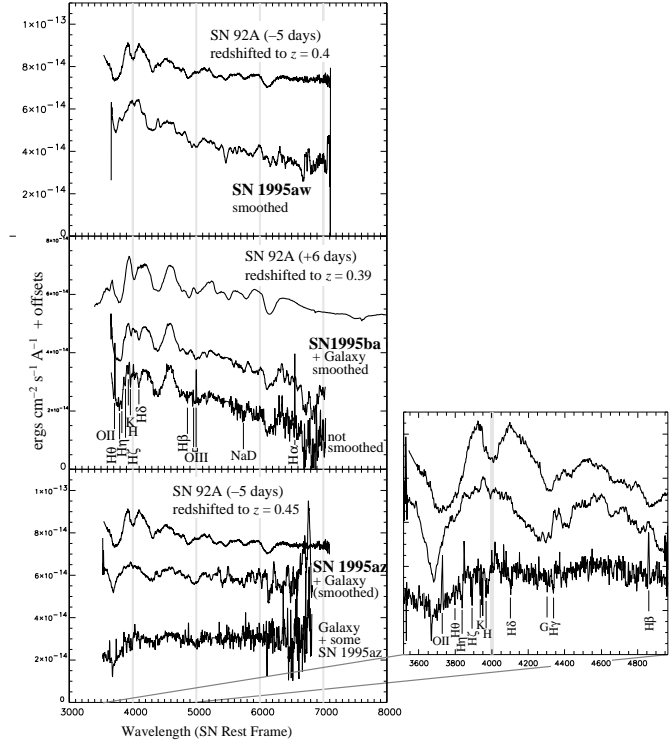}
\caption{ Spectra of SNe Ia at high z.  The use of SNe Ia 
 for deriving the cosmological parameters is improved by comparing
 distant SNe Ia spectra and those from
 SNe Ia in a low z sample (see Perlmutter et al.
 1996; Nugent et al. 1996).}  
\end{figure}

\bigskip  
    
\noindent
       The first attempts to discover SNe Ia at large redshift were
   made in the searches by Noergaard--Nielsen et al. (1989) 
  using the ESO 1m telescope. 
   They resulted in the discovery of only two supernovae at high redshift
   in several years. 
   A step beyond was finally achieved by the high--z supernova search
   by Perlmutter et al. 1996 (also called Supernova Cosmology Project) 
  which started to discover 
   dozens of supernovae
   at high redshift in discovery periods of only a few days. The Supernova
   Cosmology Project operates at telescopes of a number of observatories 
   including ESO, CTIO, and La Palma Observatory, among others.

\bigskip

\noindent
    Other groups have started as well high--z SNe Ia searches: 
   The High--z SN Search  (Schmidt et al. 1996) 
   operates at the CTIO,
   Mount Stromlo and Siding Spring, and Kitt Peak observatories.  
   The Abell supernova search uses the  southern hemisphere telescopes
   to discover SNe Ia in Abell clusters. 

\bigskip
   
\noindent
       Follow--up of light curves and spectra (see Figure 3) 
  is done and the decline--brightness 
   relationship is used in the construction of the Hubble high--redhift 
   diagram. Goobar \& Perlmutter (1995) have reformulated 
   the goal of determining
   the deceleration parameter by showing that it is possible to determine
   simulatenously $\Omega_{M}$ and $\Omega_{\Lambda}$. Thus, the goal is
   to measure the matter density of the Universe and decide whether a 
   non--zero value of the cosmological constant is at work in our Universe.   
  This is done through the separate contributions of both factors
   to the luminosity distance, $\rm d_{L}$ (see Figures 4 and 5): 

 \begin{equation}
 m(z) = M + 5\ log\ d_{L} (z, \Omega_{M}, \Omega_{\Lambda}) - 5\ log\ H_{0} +
 K_{c} + 25
 \end{equation}

\begin{figure}
\input epsf2
\epsfxsize=250pt
\epsfbox{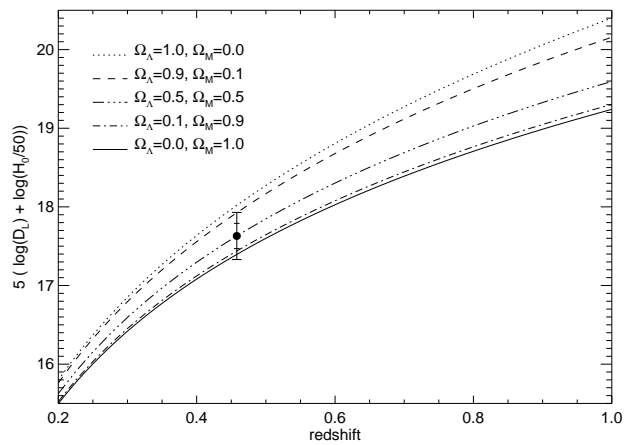}
\caption{The m(z) relation for different choices of
 $\Omega_{M}$ and $\Omega_{\Lambda}$
 according to Goobar \& Perlmutter (1995). New values will allow 
 to discriminate among the possibilities.}
\end{figure}

\begin{figure}
\input epsf2
\epsfxsize=250pt
\epsfbox{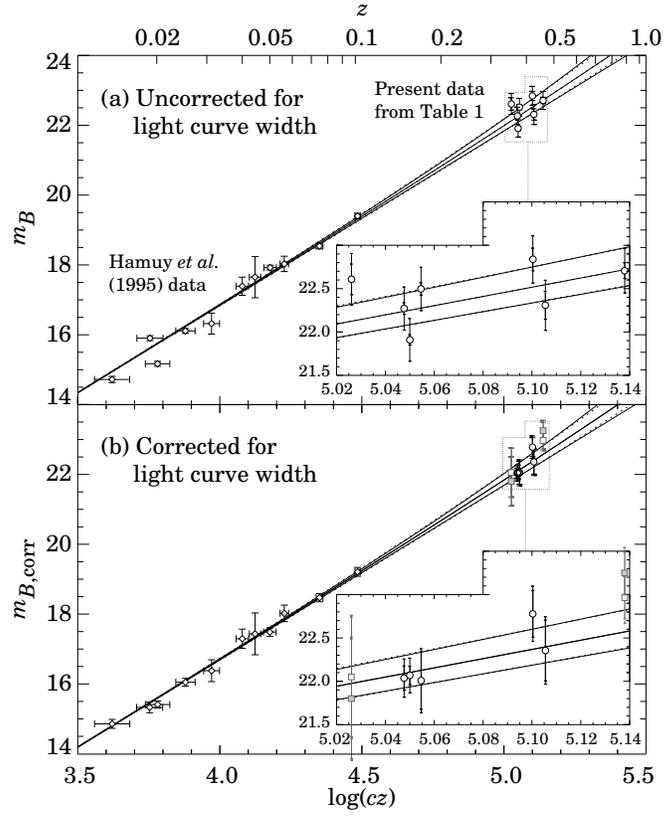}
\caption{
    In this figure it is shown the 
   the effect of the brigthness--rate of decline relationship 
   in the final diagram m(z) for the early high--redshift SNe Ia
   sample (Perlmutter et al. 1996).}
\end{figure}

\section{Number counts of SNe Ia and the geometry of the Universe} 

\smallskip

  Another way to approach the evaluation of the
  geometry of the Universe is through the number counts of SNe Ia. 
  The method has recently been proposed 
  by Ruiz--Lapuente \& Canal (1997), and the uncertainties and future
  perspectives are discussed in that work.  
  Its basis is to predict the number counts of SNe Ia in different
  models of Universe. The main uncertainty there is the sort of 
  binary scenario giving rise to Type Ia explosions: double degenerate
  systems  where the most massive WD accretes matter from the disrupted 
  WD companion, or  single degenerate systems where 
  a WD accretes from a 
  main--sequence star or a giant star. The expected number of Type Ia 
  explosions from 
  diverse scenarios are very different depending on the type of
  Universe we are in (according to the various possibilities for 
  $\Omega_{M}$ and $\Omega_{\Lambda}$), 
  and their  evolution 
  going towards earlier cosmic times clearly points out to one scenario
  or another. 
    This approach can be used to determine both the sort of explosion
   mechanism for SNe Ia and the sort of Universe in which they explode. 
  The advantage of using number counts of SNe Ia in relation to 
  that of other objects is that the luminosity corrections to apply 
  are very limited. Uncertainties mainly come from the star formation
  history and the particular evolutionary uncertainties in each 
  framework for explosion. 

\bigskip

\noindent
   The first steps towards establishing a global star formation history in 
  the Universe have been done observationally. A peak of star formation
  at z=2 has been detected and the shape of this function starts to
  become available (Madau 1996; Madau et al. 1996). 
   On the other hand, the paths towards explosion in different scenarios
  have been studied and the calculations performed by various  authors
  result in agreement on the expected behavior. The hypotheses for each
  evolutionary path can be contrasted with observations of the
  precursor objects of the finally exploding WDs, such as 
  planetary nebulae, mass accreting X--ray sources etc. 
 (Ruiz--Lapuente \& Canal  1997).
    
\bigskip

\noindent
   On the observational side, searches for rates of SNe at high redshifts are 
  providing the
  first statistical determinations of the number of SNe exploding as a
  function of magnitude: SNe Ia year$^{-1}$ deg$^{-2}$ mag $^{-1}$ (see 
results by Pain et al. 1996).
   Searches conducted at different observatories will 
   very soon provide new results.

\begin{figure}
\input epsf2
\epsfxsize=250pt
\epsfbox{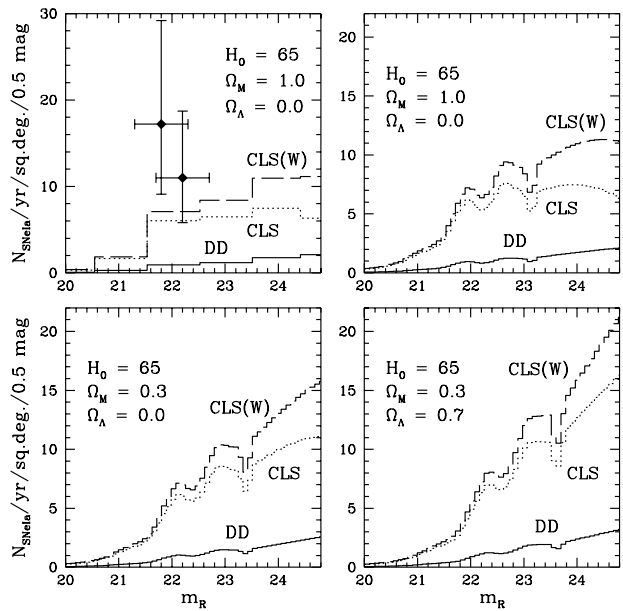}
\caption{The use of SNe Ia counts to determine
 the geometry of the Universe, from
 Ruiz--Lapuente \& Canal 1997. DD stands for the double degenerate scenario
 for SNe Ia and CLS for the most favorable single degenerate one.
 The observational points are from
 Pain et al. 1996, and Pain 1997, private communication.}
\end{figure}

\bigskip

\noindent
   The method is different from the one using the luminosity of SNe Ia 
    to derive  $\Omega_{M}$ and $\Omega_{\Lambda}$. The
   number density of explosions taking place N(z) will be an indication of the 
   geometry of the Universe instead of the observed brightness along redshift
   m(z). 
   Understanding the evolution in number of the SNe Ia exploding 
  along redshift 
  space would provide as well a step ahead to approach the chemical 
  history of the Universe in a more certain way.

\section{ Contrasting values}

We gave a look above to the values of H$_{0}$ derived by 
various authors. We would like to point out some of the results
obtained so far through the use of Type Ia supernovae in 
determining $\Omega_{M}$ and $\Omega_{\Lambda}$. The
last available values point towards a Universe with significant 
matter density (Perlmutter et al. 1996) and q$_{0}$ positive. 
 However, the error bars are 
still large.
Due to the fact that the early sample was clustered at z=0.4 it has not 
been possible to discriminate independently $\Omega_{M}$
 and $\Omega_{\Lambda}$ (see Figure 7). 
The preliminary observations give an Universe with
$\Omega = 0.88^{+0.69}_{-0.60}$ (for $\Lambda=0$).

\begin{figure}
\input epsf2
\epsfxsize=250pt
\epsfbox{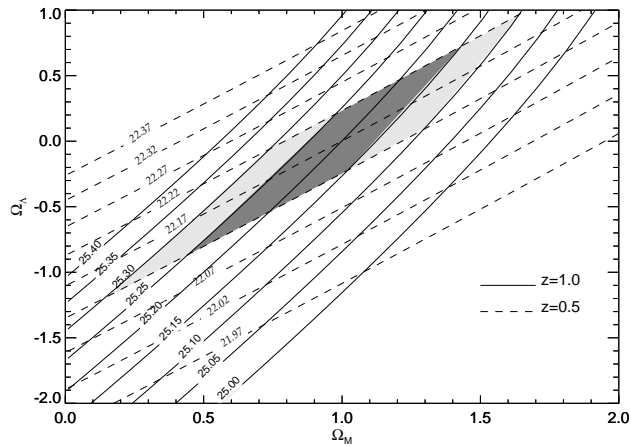}
\caption{ $\Omega_{M}$--$\Omega_{\Lambda}$ space of possible values
 according to the observations, from Goobar \& Perlmutter (1995).}
\end{figure}

\section{Measuring cosmic flows}

 Given the validity of SNe Ia as accurate distance indicators, they
 can be used to trace bulk motions in the Universe. 
  Riess, Press \& Kirshner (1995b) using the light curve shapes 
 approach, investigated  peculiar motions at moderate redshift.
 Analyzing the distribution on the sky of velocity residuals from a 
 pure Hubble flow for 13 SNe Ia they found the best solution for 
 the motion of the Local Group to be of 600 $\pm$ 350 km s$^{-1}$ 
 in the direction $\it l$ = 260$^{0}$, $\it b$ = +54$^{0}$ (see Figure 8).
 This illustrates the power contained in a sample of accurate light curve
 measurements to constrain cosmic flows.

\smallskip

\begin{figure}
\input epsf2
\epsfxsize=250pt
\epsfbox{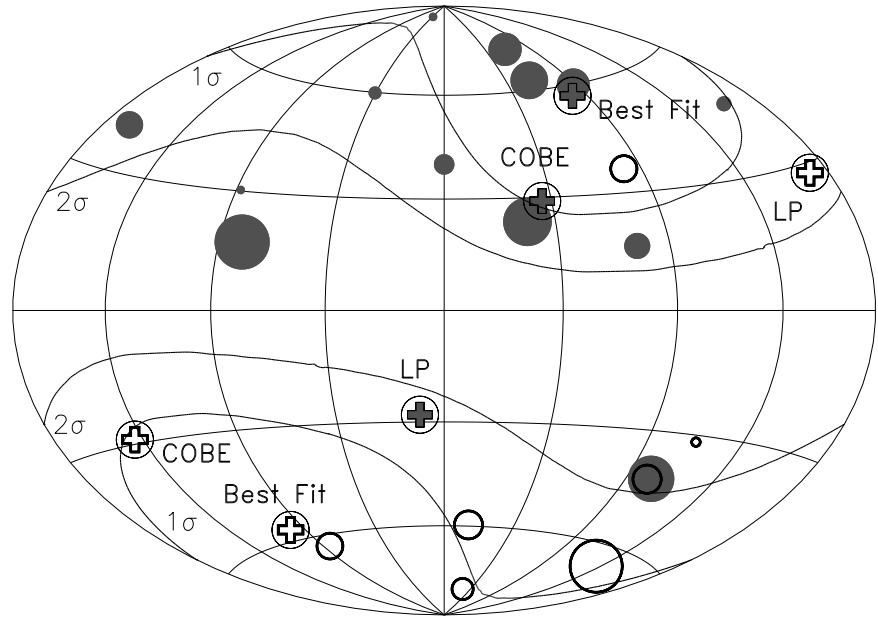}
\caption{Derivation of peculiar motions 
from SNe Ia by Riess, Press \&
 Kirshner (1995b). 
 Filled/open crosss show the direction toward which the Local Group 
 is approaching/receding according to the best fit for their SNe Ia data.}
\end{figure}

\section{The age of the Universe, the matter density and the Hubble constant}

The age of globular clusters place a constraint to the age of the 
Universe of 14 $\pm$ 2 Gy (Chaboyer 1995; Jim\'enez 1997). 
For $H_{0}$ of 60--70 and  $\Omega$ larger 
than 0.3, the age of the Universe according to  
Friedmann cosmologies is shorter than 12 Gy. For $\Omega$ = 1 and  
 $H_{0}$ in the previous range of values, the age is shorter than 10 Gy.
A non--zero cosmological 
constant  $\Lambda$ as first introduced by Einstein (1917), could give
 an answer to the
contradiction between observed age values and the expected range of ages of 
 universes within the favored values of the cosmological parameters.
But the value of $\Lambda$ is also limited by SNe Ia observations
(so far $q_{0}$ seems to be positive). 

\bigskip

\noindent
   If, as suggested by Tammann and Sandage (1995), 
 $H_{0}$ is around 50, the age of the
 Universe is comfortably larger and neither a non--zero cosmological
 constant, nor deviations from the Friedmann universes
 are needed. 

\bigskip

\noindent
 However, if a global value of $H_{0}$ around 65 is confirmed, and 
 the Universe is flat with q$_{0}$ close to 0.5, an 
 alternative to the well tracked classical paths should be sought. 
 Dabrowski \& Hendry (1997), for instance, have made
 a reanalysis of the situation and 
 show that in inhomogeneous universes the above values of 
  $H_{0}$ and  q$_{0}$ imply ages 
 largely compatible with the limits on the age of the Universe 
 resulting from the globular cluster ages. 
 If a value of $H_{0}$ between 60 and 70 is confirmed, and the Universe
 is as dense as suggested by the preliminary Perlmutter et al. (1996) 
 results,  the door to old and new alternatives 
 would reopen again.

\acknowledgements
 I would like to thank the organizers of the Casablanca
 Winter School for inviting me to lecture on cosmological uses of 
 supernovae. I thank as well A. Riess and S. Perlmutter for
 the permission to reproduce figures from their work. 
 Research on  cosmology through supernovae
 by the author is supported by the Spanish DGYCIT.

\end{document}